\documentstyle[12pt,aaspp4,flushrt]{article}
\slugcomment{Accepted 7 May 1998 by The Astrophysical Journal}
\lefthead{Alves et al.}
\righthead{Dust Extinction and Cloud Structure: L977}

\def\msun{M$_\odot$\ }

\def\h2{H$_2$\ }

\def\pc3{pc$^{-3}$}

\def\h {$^h$}

\def\deg{$^{\circ}$}

\def\am{$^{\prime}$}
\def\as{$^{\prime\prime}$}
\def\ltsima{$\;\buildrel<\over\sim\;$} 
\def\simlt{\lower.5ex\hbox{\ltsima}}
\def\gtsima{$\;\buildrel>\over\sim\;$} 
\def\simgt{\lower.5ex\hbox{\gtsima}}
\def\cao{\c{c}\~{a}o}

\def\ii{\'{\i}}

\begin{document}

\vskip 1.1cm
\title{Dust Extinction and Molecular Cloud Structure: L977}

\vskip 0.2cm
\author{Jo\~ao  Alves\altaffilmark{1,2}, Charles J. Lada
\altaffilmark{1}}
\affil{Harvard-Smithsonian Center for Astrophysics, 
60 Garden St., Cambridge MA 02138}
\author{Elizabeth A. Lada\altaffilmark{1}}
\affil{Astronomy Department, University of Florida,
Gainesville, FL 32608}
\author{Scott J. Kenyon\altaffilmark{1}}
\affil{Harvard-Smithsonian Center for Astrophysics, 
60 Garden St., Cambridge MA 02138}
\author{Randy Phelps\altaffilmark{1}}
\affil{Observatories of the Carnegie Institution of Washington,
813 Santa Barbara Street, Pasadena, CA 91101}

\authoremail{jalves@cfa.harvard.edu}

\altaffiltext{1}{Visiting Astronomer, Kitt National Observatory, part of the
National Optical Astronomy Observatories, which is operated by the Association
of Universities for Research in Astronomy, Inc. under contract with the
National Science Foundation.}
\altaffiltext{2}{Also Physics Department, University of Lisbon, Lisbon Portugal}

\begin{abstract}

We report results of a near--infrared imaging survey of L977, a dark
cloud in Cygnus seen in projection against the plane of the Milky Way.
We use measurements of the near--infrared color excess and positions
of the 1628 brightest stars in our survey to measure directly dust
extinction through the cloud following the method described by Lada et
al.  (1994).  We spatially convolve the individual extinction
measurements with a square filter 90$^{\prime\prime}$ in size to
construct a large--scale map of extinction in the cloud.  We integrate
over this map to derive a total mass of $M_{L977} = (660 \pm 30) \
(D/500 pc)^2$ M$_\odot$ and, via a comparison of source counts with
predictions of a galactic model, estimate a distance to L977 of 500
$\pm$ 100 pc.  We find a correlation between the measured dispersion
in our extinction determinations and the extinction which is very
similar to that found for the dark cloud IC 5146 in a previous study.
We interpret this as evidence for the presence of structure on scales
smaller than the 90$^{\prime\prime}$ resolution of our extinction map.

To further investigate the structure of the cloud we construct the
frequency distribution of the 1628 individual extinction measurements
in the L977 cloud.  The shape of the distribution is similar to that
of the IC 5146 cloud.  Monte Carlo modeling of this distribution
suggests that between 2 $<$ A$_V$ $<$ 40 mag (or roughly 1 $<$ $r$ $<$
0.1 pc) the material inside L977 is characterized by a density profile
$\rho \,(r) \propto r^{-2}$.  Direct measurement of the radial profile
of a portion of the cloud confirms this result.

At the lower galactic latitude of L977, we find both the mean and
dispersion of the infrared colors of field stars to be larger than
observed toward IC 5146.  This produces an increase of about a factor
of 2 in the minimum or threshold value of extinction that can be
reliably measured toward L977 with this technique.  Nevertheless the
accuracy in an extinction map pixel is not significantly different
toward L977 due to the increased number of field stars at this
latitude.  We also find an increase in the number of detected giant
stars at the lower galactic latitude of the survey by almost a factor of
two.  Most of these excess stars suffer extraneous extinction and are
probably red giants seen along the disc of the Milky Way up to
distances $\sim$ 15 kpc and reddened by unrelated background molecular
clouds along this direction of the Galaxy.  We discuss a possible
application of this observable to galactic structure studies on the
plane of the Galaxy.

\end{abstract}


\section{Introduction}

Knowledge of the distribution of mass within molecular clouds is
critical for the understanding of how molecular clouds form, evolve,
and ultimately produce stars and planets.  Dust is the most reliable
tracer of the undetectable molecular hydrogen gas, which is the
primary component of the mass of molecular clouds.  The most
straightforward way to measure the dust content of a molecular cloud
is through direct measurements of dust extinction of background
starlight.  This can be optimally done at near infrared wavelengths
where the dust opacity is modest and the reddening law appears
universal (Jones \& Hyland 1980; Cardelli et al.  1989; Martin \&
Whittet 1990; Whittet et al.  1993).  The determination of gas mass
from dust extinction is relatively straightforward and relies on
basically one observationally well established assumption:  a constant
gas--to--dust ratio (Lilley 1955; Jenkins \& Savage 1974; Bohlin et
al.  1978).

Recently, Lada et al.  (1994) demonstrated that large scale
near-infrared imaging surveys of dark clouds can directly trace the
distribution of dust extinction and column density through a molecular
cloud at considerably higher angular resolution and greater optical
depth than previously possible from pure star counts (Bok 1937).  In
fact, Lada et al.  (1994) developed a method for measuring and mapping
extinction which is more direct and fundamentally more powerful than
star counts made at any wavelength.  Their method uses:  1)
measurements of infrared (H[1.6$\mu$m]$-$K[2.2$\mu$m]) color excess,
2) knowledge of the near--infrared extinction law, and 3) techniques
of star counting to derive directly and map extinction through a
cloud.  Also, since the method is based on pencil beam measurements of
the extinction, it can give information about both the large--scale
and small--scale structure of a molecular cloud.

We have two main motivations for the present paper.  First, we want to
apply the Near--Infrared Color Excess method used by Lada et al.
(1994) (hereafter, NICE method) to a different molecular cloud to
compare its physical conditions with those derived for IC 5146.
Second, because the information on the spatial distribution of dust
derived from this method is a function of both the surface density and
colors of stars detected through the target cloud, we want to test the
applicability of the method against rich background star fields at low
galactic latitudes.

L977 \footnote {Two dark clouds from the updated Lynds catalogue
(Lynds 1962; Nagy 1979) fall inside the Near--Infrared (NIR) survey
presented in this paper:  L977 and L981.  We choose L977 as
the representative cloud since it has the largest  area (0.200 vs.
0.019 deg$^2$).  In the original version of the catalogue this cloud
has the number 978.}  is part of dark globular filament GF7 in the
Cygnus complex (Schneider \& Elmegreen 1978) and lies against a rich
background of field stars towards the warped plane of the Galaxy ($l =
90$\deg, $b = 2$\deg).  There is no obvious sign of ongoing star
formation in this cloud.  Although three IRAS point sources lie in the
direction of L977, their colors are indicative of infrared cirrus
emission rather than pre--main sequence objects.  This dark cloud is
in the Dobashi et al.  (1994) $^{13}$CO large scale survey of
the Cygnus region as an isolated cloud and free from confusion
with other molecular clouds.  Apart from this survey L977 is virtually
unstudied.

We describe the acquisition and reduction of the observations in
Section 2,  the results in Sec. 3,  the analysis and discussion
in Sec.  4, and we summarize our conclusions in Sec. 5 of the paper.

\section{Observations and Data Reduction}

We obtained near-infrared imaging observations of L977 using the
National Optical Astronomy Observatory (NOAO) Simultaneous Quad
Infrared Imaging Device (SQIID) on the Kitt Peak National Observatory
(KPNO) 1.3m telescope in 1994 October.  SQIID is equipped with four
256 $\times$ 256 platinum silicide (PtSi) focal plane arrays.
Dichroic mirrors allow simultaneous observations at four infrared
wavelength bands:  $J$ (1.25 $\mu$m), $H$ (1.65 $\mu$m), $K$ (2.20
$\mu$m), and $L$ (3.4 $\mu$m).  For this project, we obtained data
only in the $J$, $H$, and $K$ bands.  The optics provide a field of
view of approximately 5\am.5 $\times$ 5\am.5 and a resolution of
1$^{\prime\prime}$.36 per pixel at $K-$band.

We observed a total of 17 fields at each wavelength band towards L977
covering 336 arcmin$^2$ on the sky in a roughly 4 $\times$ 4 mosaic
grid.  The area surveyed in the 3 NIR bands is represented in Figure 1
as the central square overlaid on the Digitized Sky Survey red POSS
plate \footnote{Based on photographic data of the National Geographic
Society -- Palomar Geographic Society to the California Institute of
Technology.  The plates were processed into the present compressed
digital form with their permission.  The Digitized Sky Survey was
produced at the Space Telescope Science Institute under US Government
grant NAG W-2166.}.  The L977 molecular cloud is readily seen as the
zone of obscuration against the rich star field that characterizes
this region of the galaxy.  The fields observed in the direction of
our target cloud were spatially overlapped by $\sim$1\am.5 in both
right ascension and declination, allowing for both the accurate
positional placement of the mosaicked fields and redundancy of our
photometric measurements of sources located in the overlapped regions.

To characterize the distribution of background/foreground field stars
a total of 7 control fields were observed at each band.  Since our two
target clouds lie against a complicated background of field stars, very
near the galactic plane, special care was used in choosing the control
fields regions.  These fields were chosen to be as close as possible
to our target cloud, equally distributed around it, but free from any
significant molecular cloud material contamination through inspection
of the $^{13}$CO survey (Dobashi et al.  1994) and the Palomar Sky
Survey Prints.  The boxes to the East and to the West of L977 in
Figure 1 show the regions where these control fields were taken.
They are spatially separated by 1 degree in the sky ($\Delta l =
0.7$\deg, and $\Delta b = 0.8$\deg).  The control field images were
not overlapped.  The integration time in each filter for observations
both on and off L977 was 180 seconds.  All fields were observed twice,
with a 15\as\ offset between observations.

\subsection{Reduction Procedure}

We used a combination of the standard Image Reduction and Analysis
Facility (IRAF) routines and custom software for data reduction.  The
data frames were reduced in the following 3 step procedure:  a dark
frame was subtracted from each data frame, the resulting
dark--subtracted frame was then divided by a flat field, and finally
the value of the local sky was subtracted from the dark subtracted
flat fielded frame.  The dark frames are averages of 20 individual
dark images acquired at the beginning and end of each night.  The flat
fields used for each filter were constructed by median filtering the
20 time-neighboring (at roughly the same airmass) dark subtracted images
and found to be indistinguishable within the errors.  These flat-field
images were then autonormalized before division into the appropriate
dark subtracted data frames.  The local sky was taken as the mean of
the dark--subtracted flat--fielded data frame.  A conventional sky
removal procedure (direct frame--by--frame sky subtraction) was
performed on 6 typical data frames for comparison with the adopted
reduction procedure.  The photometric results from both image
reduction procedures were then compared and found to be identical
within the photometric errors discussed below.

\subsection{Source Extraction and Photometry}

Sources were identified and counted using the DAOFIND routine (Stetson
1987) within the IRAF package.  DAOFIND was run on each image
separately, using a full width at half maximum (FWHM) of the point
spread function between 1.7 to 2.0 pixels and a single pixel finding
threshold equal to 5 times the mean noise of each image.  The results
and images were carefully inspected.  The DAOFIND coordinate files
were edited to remove bad pixels, non-stellar objects and artifacts
misidentified as stars and to append stellar sources that were not
originally extracted by the finding routine.  Photometry was then
obtained for all extracted stars in the final images using APPHOT with
a variety of different apertures.  Photometric measurements were
independently obtained for each dithered pair of images toward each
field.  Custom stellar matching software was used to compare the
magnitudes obtained for each dithered stellar pair.  A series of tests
were then performed on this photometric database.  For the various
apertures, plots of the difference magnitudes for each star versus the
mean magnitude showed no systematic offset, while still displaying the
expected flared errors at faint magnitudes.  Further, plots of the
difference magnitude versus the internal photometric uncertainties,
returned by APPHOT, showed no unexpected effects or features.  We
conclude that the internal APPHOT uncertainties are reasonably good
estimators of the true photometric uncertainties.  In Figure 2 we plot
the photometric errors, as returned by APPHOT, in the colors
($(J-H)$ open triangles; $(H-K)$ filled triangles) as a function of
magnitude $(H)$ for stars observed toward L977.

For the photometry presented here, the 2.5 pixels (3\as.4) aperture
was chosen.  This aperture size corresponds to $\sim$ 2--3 times the
FWHM of the typical point spread function for our SQIID data.  Sky
values around each source were determined from the mode of pixel
intensities in an annulus having an inner radius of 10 pixels and an
outer radius of 20 pixels.  The average number of stars found per
frame was about 200 at K, reducing the need for PSF fitting
photometric extractions.  Using an aperture with a 2.5 pixel radius
insured that the flux from $\sim$90\% of the sources was not
contaminated by the flux from nearby stars.  However, this aperture
was too small to contain all the flux from a source.  Therefore, to
account for the missing flux, multi-aperture photometry was performed
for each image on all bright (m$_{J,H,K}$ $<$ 12), isolated sources.
For these sources, the flux obtained in the 2.5 pixel radius aperture
was compared with flux measured in a larger 3, 4, 5, 7, and 9 pixels
radius apertures and the fraction of the total source flux falling
within the 2.5 pixel radius aperture was then determined.  Typically,
the 2.5 pixel radius aperture was found to contain approximately 90\%
of the total source flux.  A similar result was reached by Barsony et
al.  (1997) for SQIID observations of the $\rho$ Ophiuchi cloud core.
The instrumental magnitudes for all extracted SQIID sources were
corrected to account for the missing flux.

We confirm the tendency for stars imaged onto the outer 10\% of the
PtSi SQIID arrays to show unusual colors (e.g., Lada et al.  1994).
To remove this effect, stars imaged onto the outer 10\% of the array
had their corresponding photometry rejected.  All surviving data were
then merged into two databases:  the target and the control database.
Duplicate stars from neighboring target frames (up to four), were
identified and their magnitudes averaged.

Photometric calibration for our data was accomplished using the list
of Elias standard sources (Elias et al.  1982).  Eight standards were
observed on the same nights and through similar airmasses as were
observations of on field and off field images and were used to
independently establish the photometric zero points for each of the
filters.  Otherwise, we elected to use the SQIID camera filters and
arrays as our photometric system and have applied no further
adjustments (e.g., color-corrections) to our measurements since the
SQIID photometric system seems close to the Elias et al.  (1982)
photometric system (CIT) (Kenyon, Lada, \& Barsony 1998).  The
standards were observed at nearly the same airmasses as the L977
fields and since atmospheric infrared extinctions are very small 
(\ltsima 0.1 magnitudes per airmass), our reported magnitudes and 
colors are likely not in error by much more than their photometric 
uncertainties determined by APPHOT.

\subsection{Completeness Limits}

In order to estimate the completeness limits of our observations, we
compared the number of stars identified, as a function of magnitude,
on pairs of dithered images in each band.  DAOFIND and PHOT were run
on the overlapping regions of 5 pairs of dithered off field images,
using the same criteria described in 2.3.  The apparent magnitude
distribution of stars that were identified and matched across the
dithered pairs of images was compared to the magnitude distribution of
stars that were found in only one of the image pairs.  From this
comparison, we estimate that the identification of sources in the
individual image frames is $\sim$ 90\% complete to m$_J$ = 15.5, m$_H$
= 14.5 and m$_K$ = 13.5.  These completeness limits are extended
another 0.4 magnitudes after combining the paired images obtained in
each on and off position.  For the $K-$band observations we also
estimated the survey completeness to fainter magnitudes and found it
to be roughly 85\% between 13.5--14 mag, 75\% between 14--14.5
magnitudes and only 60\% between 14.5--15.0 magnitudes for the
individual dithered images.

\subsection{Positions}

We determined absolute $\alpha$ and $\delta$ positions for each
individual source identified by our near-infrared observations of
L977.  We used a centering algorithm in APPHOT to obtain center
positions for the sources in pixels.  Duplicate stars on adjacent
frames were used to register the relative pixel positions of the
individual SQIID frames onto a single positional grid.  Pixel
positions were converted to equatorial coordinates using observations
of 3 stars from Hubble Space Telescope Guide Stars Catalog (GSC) that
were detected within our grid.  Both the plate scale (1.358 arcsec
pixel$^{-1}$ at $K-$band) and grid orientation relative to north
(+0.71\deg) were derived from matching the positions of these stars
with their pixel coordinates.  The resulting coordinate transformation
was applied to all stars in the image to obtain their positions in the
GSC astrometric system.  We re--measured the position of the GSC stars
and determined an average difference to the absolute positions of
0.5\as.  The measured standard deviation of the differences between
the derived and absolute coordinates of the reference stars was found
to be 0.2\as.

\section{Results}
\subsection{Color--Color Diagrams}

The results of the NIR photometry for the control regions (defined in
Figure 1 by the boxes to the East and to the West of L977) are
presented in Figure 3.  These plots display only those stars in each
region for which photometric uncertainties in each band were less than
0.10 magnitudes.  Also plotted as a solid line on both color--color
diagrams is the locus of points corresponding to unreddened main
sequence and giant stars (Koornneef 1983).  The two parallel dashed
lines define the reddening band for both main sequence and giant stars
using the reddening law of Rieke \& Lebofsky (1985).  The typical
photometric error for both colors for a $K=14$ star --- roughly a 10
sigma detection --- is displayed in the upper corner of both diagrams.
There are a total of 303 sources detected in the East control field
region and 219 in the West control field region.  The surface density
of detected sources in the $JHK-$bands for the East control field
region ($\Sigma_E$) is slightly higher than in the West control field
region ($\Sigma_W$) --- $2.6$ vs.  $2.5$ stars ($\sigma < 0.1$) per
square arcmin --- most likely due to the fact that the East control
field region is almost 1\deg\ closer to the Galactic plane.  The mean
$(H-K)$ and $(J-H)$ colors and dispersions for the two regions are
nevertheless the same to 2 significant figures:

\[<(H-K)>_E \:\, = \:\, <(H-K)>_W \:\, = 0.20 \pm 0.13 \:\, \]

\[<(J-H)>_E \:\, = \:\, <(J-H)>_W \:\, = 0.56 \pm 0.27 \:\, \]

\noindent The spatial distribution of sources in the color--color
plane is apparently very similar for both control regions.  We used a
2D version of the Kolmogorov--Smirnov test (Peacock 1983; Fasano \&
Franceschini 1987; Press et al.  1992) to compare both East and West
control fields color distributions and we found a high probability
(0.74) that the sources in both control fields were randomly drawn
from the same population.  Sources in both the East and West control
fields are distributed in two distinct groups:  1) a blue group that
is roughly uniformly distributed around the point ($(H-K)$, $(J-H)$)
$=$ (0.07, 0.35) corresponding to the color of an mid G main sequence
star (Koornneef 1983), and 2) a red group which has an elongated
distribution along the direction of the reddening vector and has
colors ($(H-K)$, $(J-H)$) $=$ (0.20 to 0.40, 0.60 to 1.10)
corresponding to the colors of either late type giants or reddened
early K giants/main sequence stars.

In Figure 4 we show the $JHK$ color--color diagram for the L977 field
defined in Figure 1 by the central square region.  The main sequence
and giant locus, as well as the photometric errors are the same as in
Figure 3.  The displacement in this diagram caused by 5 visual
magnitudes of extinction is represented by the solid line to the left
of the reddening vector.  There are 809 sources simultaneously
detected in the $JHK-$bands with photometric errors below 0.10
magnitudes at each band.  The large majority of sources is distributed
along the reddening vector.  Of the 809 high--quality photometry
sources shown, 84 sources are located below the reddening band and 79
sources are located above it.  Hence, the number of infrared excess
sources detected toward the L977 field is negligible.  The ten most
deviant sources from the reddening band are all faint and have
magnitudes very close to the 10 sigma detection limit.  The average K
magnitude of this sub-sample of 10 stars is $< K >$ $=$ 13.60 $\pm$
0.26 magnitudes.  Comparison of Figures 3 and 4 strongly suggests that
the vast majority of stars observed toward the L977 molecular cloud
are field stars unrelated to the cloud.  Moreover, a significant
number of these sources are clearly reddened background objects
observed through the cloud.  Because the surveyed area (Figure 1)
included a substantial region in the sky suffering negligible
extinction from L977 we are able to retrieve the results obtained in
the control field regions, namely the blue and red grouping of sources
with the same characteristics and locations on the color--color plane.

\subsection{Determination of the NIR Extinction Law}

Although we have assumed the extinction law of Rieke \& Lebofsky (1985)
for comparison with our data, we can use our observations to derive
the NIR extinction law, $E(J-H)/E(H-K)$, in this cloud.  This can be
done from comparison between the color of stars seen through L977 and
color of stars in the control fields.  We have to assume that the
background population to L977 is well characterized by the stars in
the control fields.  As we shall see in section 4.1, that is a valid
assumption.  However, the dispersion in the color of stars in the
control fields at these low galactic latitudes makes a photometric
determination of the extinction law uncertain.  Because we have a
fairly large sample of stars in the L977 field we can improve on this
uncertainty by creating a sub--sample of reddened stars seen through
L977 for which we can claim a better knowledge of their intrinsic
properties, hence, a more certain estimation of their color excesses.
The obvious choice, for their intrinsic brightness, is a sub--sample
of reddened giant stars seen through L977.  We create this sub--sample
by excluding from the data presented in Figure 4 stars with:  i)
$(J-H)<1.3 \,\,\, \wedge \,\,\, (H-K)<0.5$ mag, and ii) $K>11$ mag.
The first condition assures that all the stars in this sub--sample
have colors that are redder than the colors of stars in the control
fields (Figure 3), hence background to L977.  The second condition
assures that most of, if not all, the stars in the sub--sample are
giant stars; in fact, the large majority of main--sequence stars at
distances greater than the distance to this cloud (500pc, see section
4.2) and behind a curtain of A$_V$\gtsima 5 mag --- to meet condition
i) --- have $K>11$ mag.

The 26 stars that constitute the sub--sample 
have highly correlated $(J-H), (H-K)$ colors, with a Spearman
rank--order correlation coefficient of $r_s = 0.90$ (the value
of this coefficient before application of condition ii) was 0.74).
To determine the color excesses of stars in the sub--sample 
we assume that their intrinsic color is given by the average color
of stars in the red group of the control fields (Figure 3) with 
$K<11$ mag. A linear fit to the derived color excesses gives:

\[ E(J-H)/E(H-K) = 1.6 \pm 0.1 \]
 
\noindent where 0.1 is the statistical uncertainty.  The systematic
uncertainty caused by the rms dispersion in the assumed intrinsic
color is, nevertheless, $\sim$0.2.  The derived extinction law is
consistent within the uncertainties with previous results (see for a
review Kenyon, Lada, \& Barsony 1998) but the large errors caused by
the complicated stellar background preclude a more detailed
comparison.  Throughout this paper we will still use the Rieke \&
Lebofsky (1985) extinction law, $(E(J-H)/E(H-K))=1.7$, to facilitate
comparison of our results with previous work.

\section{Analysis and Discussion}

\subsection{Application of the NICE Extinction Determination}

The line--of--sight extinction to an individual star can be directly
determined from knowledge of its color excess and the extinction law.
The infrared color excess is proportional to dust extinction and thus
column density and can be directly derived from the observations
through:

\[ E(H-K)\  = \ (H-K)_{observed} \ - \  (H-K)_{intrinsic} \]

\noindent The intrinsic $(H-K)$ color of main--sequence and giant
stars have a narrow enough range (Koornneef 1983, Bessel \& Brett
1992) that even without further information a reasonable determination
(1$\sigma \sim 2.5$ mag of visual extinction) of the line--of--sight
extinction (e.g., A$_K$) towards any significantly extincted star
can be made.  This determination can be considerably improved
through observations of control fields around the target cloud since
we expect the colors of field stars to be dominated primarily by
K--giants, assuming the differential reddening is small.  The infrared
color excesses derived in this way can then be converted to an
extinction A$_\lambda$ at a given wavelength $\lambda$, via the
extinction law.

We start with the assumption that stars observed in our off cloud
control fields can serve as a surrogate for the stellar population
directly behind the cloud.  To test the validity of this assumption we
can search for possible color variations across the background by
comparing our control fields on both sides of L977.  The positions of
our control fields were carefully selected to be close to the
molecular cloud but distant enough to be free from contamination by it
(see Figure 1).  Although the two fields are 1\deg\ apart in the sky
(and almost 1\deg\ apart in Galactic latitude) our photometry (Figure
3) shows no sign of a color variation.  Both fields have virtually the
same mean $(J-H)$ and $(H-K)$ colors, dispersions, and similar spatial
distributions on the color--color plane.  The angular size scale for
color variations on the near-infrared colors of field stars must then
be greater than 1\deg\ in this direction of the Galaxy and we can
assume with a high level of confidence that the $(H-K)$ colors of
stars behind the cloud are well represented by the $(H-K)$ colors of
stars in the control fields.  The mean value of the latter can then be
taken more securely to characterize the typical intrinsic $(H-K)$
color of stars behind the cloud:

\[(H-K)_{intrinsic}\  \equiv \ <(H-K)>_{control}\  = \ 0.20 \pm
0.13 \ \rm mag \]

\noindent where the quoted uncertainty is the dispersion in $(H-K)$
control field colors.

We used the above equations to derive individual color excesses to the
1628 stars observed through L977.  We converted the individual
measurements to equivalent visual extinctions, A$_V$, using a standard
reddening law (Rieke \& Lebofsky 1985):

\[ A_V = 15.87 \times E(H-K) \]

Although the individual A$_V$'s derived in this manner are
proportional to the true dust column density along the line--of--sight
to each star they only accurately reflect the true visual extinctions
as long as the assumed reddening law is appropriate for this cloud.
Grain growth in cold clouds alters the extinction law at wavelengths
$\ll 1\mu$m.  Conversion to a near--infrared extinction (e.g., A$_K =
1.76$ E$(H-K)$) would avoid this problem since the NIR reddening law
does not vary significantly with grain growth in dark clouds (e.g.,
Mathis 1990).  However, to facilitate comparison with previous work we
follow convention and express our dust opacities in terms of
equivalent visual extinctions.

Because of dust extinction, stars that will be detected through the
increasingly denser regions of L977 can only be the brighter
background stars.  It is important then to know how the mean $(H-K)$
color and dispersion will vary with $K$ brightness as this variation
might affect the correct determination of extinction from color
excesses.  To search for such possible variation we present in Figure
5 the color--magnitude diagram for all the stars in the control
fields.  The open circles represent sources with $(J-K) < 0.5$ mag
(the blue group, described in Sec.  3.1) and the filled circles
represent sources with $(J-K) > 0.5$ mag (the red group).  To evaluate
how the mean $(H-K)$ color and dispersion vary as a function of $K$
brightness we binned the data in $K$ (2 magnitudes bins) and
determined the mean $(H-K)$ color and dispersion for each bin.  This
mean $(H-K)$ color and dispersion is represented in Figure 5 as the
open squares and the error bars, respectively.  We conclude that while
the dispersion is essentially the same for all bins there is a slight
increase in the mean $(H-K)$ color from fainter to brighter
magnitudes.  This is caused by the larger number of red sources (most
likely red background giants; see Appendix) that populate the bright
end of the color--magnitude diagram.  Nevertheless, this increase
($\sim$ 0.05) is less than half the average dispersion ($\sim$ 0.13)
and has a negligible effect on the determinations of extinction.

In Figure 6 we plot the frequency distribution of A$_V$'s derived
toward L977.  For comparison we also plot the distribution of A$_V$'s
for IC 5146 using the data of Lada et al.  (1994).  The systematic
offset between both distributions is caused by the richer background
of L977 at lower galactic latitudes.  The shape of these distributions
is a function of both the detailed distribution of matter within the
cloud and the detection limits of the infrared survey at high
extinctions.  The distributions in both clouds fall off steeply with
increasing extinction in a fashion roughly consistent with a
power--law.  This indicates that the amount of material at high
extinctions is relatively small in both these filamentary dark clouds.
At high extinctions the steepness of the distribution is enhanced by
the inability to detect heavily reddened faint stars imposed by the
sensitivity of the NIR survey.  Since the sensitivity of both surveys
is the same, the similarity of the shapes of the extinction frequency
distributions suggests that both L977 and IC 5146 are characterized by
a similar intrinsic structure, i.e., a comparable proportion of high
column density to low column density gas.  In Section 4.5 of this
paper we will model this frequency distribution for various clouds
configurations.

To characterize the large--scale global structure of the cloud we
spatially convolve the individual pencil--beam measurements (which
represent a random spatial sampling of the cloud extinction) with a
square filter 90$^{\prime\prime}$ in size and produced an ordered map
of the extinction uniformly sampled at the Nyquist frequency.  The map
is shown in Figure 7 superposed on a digitized POSS red plate of the
region.  The 1$\sigma$ confidence level on the derived extinction
measurements toward individual stars is set by the rms dispersion 
in the $(H-K)$ color of background stars measured in the control 
fields (Figure 3) and is equal to $\sim$ 2 magnitudes of visual 
extinction.  The 1$\sigma$ confidence level on a map pixel is 
nevertheless $\sim$ 1 magnitude (see discussion in Section 4.6).  
Stars suspected to be foreground to the cloud (see next
section) were removed from the database.  Contours start at 4 visual
magnitudes of extinction (4$\sigma$) and increase in steps of 2
magnitudes up to 24 magnitudes of visual extinction.  The
discontinuity near the bright star SAO 050355 is due to lack of one
SQIID field in our data.  We decided not to take this field as a
precautionary measure since this SAO star, a M3 star (Neckel 1974)
with an estimated $K=$ 3.8 mag, would potentially affect the SQIID
detector.  The overall shape of the map correlates well with the shape
of the more opaque regions of the molecular cloud (see Figure 7).

\subsection{Distance Determination}

The only reference in the literature regarding the distance towards L
977 appears in Dobashi et al.  (1994) $^{13}$CO survey.  These authors
estimate a distance of 800 pc from association to nearby astronomical
objects with known distances.  We used the Wainscoat et al.  (1992)
infrared model of the Galaxy to determine the expected number of
foreground stars seen toward the opaque regions of L977 and
independently determined a distance to this cloud.  The molecular
cloud can be used as a wall to discriminate between
foreground/background stars as foreground stars are easily detected
against the regions of high extinction.  We searched for these stars
in regions that presented an average extinction above 8 magnitudes
(the central region of the L977 filament, $\sim$ 36 arcmin$^2$) and
compared the number of stars with derived extinctions A$_V < 2$ mag (5
stars) with the expected number of foreground stars from the galactic
model.  The $K-$band surface density of stars of the control fields
was used to calibrate the galactic model.  In Figure 8 we present the
expected number of foreground stars toward the denser (A$_V >$ 8 mag)
regions of L977 as a function of distance.  We estimate a distance to
L977 molecular cloud of 500 $\pm$ 100 pc.

Extra information regarding a lower limit to the distance to L977 can
be obtained from studied bright stars clearly foreground to the cloud.
There are two such bright stars seen in projection against the
molecular cloud complex:

\begin{enumerate}
\item Less than 1\deg\ North of L977 (in front of L988, a molecular cloud
associated in projection and velocity with L977) lies SAO 033091 (HR
8072), a K0III star with a visual magnitude of m$_V = $ 6.37 (Breger
1968, Strassmeier et al.  1994).  This star was observed with the
astrometric satellite HIPPARCOS (Perryman et al.  1997) and an
accurate distance of 119$\pm$8 pc was derived from trigonometric
parallax (a spectrophotometric parallax would have given a distance
$\sim$ 150 pc).  The distance determined by HIPPARCOS can be taken as
a conservative lower limit of the distance to L977.  

\item The second bright star, not as well studied as SAO 033091, is
seen in projection against L977 (SAO 050355, see Figure 1) .  It was
classified as a M3 star by Neckel (1974) but there is no information
available on its luminosity class.  If we assume a main-sequence,
giant, or supergiant luminosity we can determine a distance of d$_V =
\,\,\sim 8$ pc, d$_{III} = \,\,\sim 580 $ pc, or d$_I = \,\,\sim 4800$
pc respectively to this star.  In any case the derived extinction is
negligible which puts this star in the foreground of L977.  We can
readily dismiss the derived distance of 4800 pc:  1) at this distance
we would be seeing this star at least through two of the Galaxy's
spiral arms ($l = 90$\deg, $b = 2$\deg) which is not compatible with
the negligible extinction measured towards this star and 2) at this
distance there would be no contrast to distinguish the shape of L977
against the background field due to foreground stars which is clearly
not the case.  The other two derived distances cannot be dismissed but
while d$_V = \,\,\sim 8$ pc would bring no further constraints to this
discussion, the one determined for the giant star case (580 pc) is
still consistent, within the uncertainties, with a distance of 500
$\pm$ 100 pc derived from the galactic model.

\end{enumerate}

The uncertainty associated with the distance determination will be the
largest source of systematic error in the derived properties of L977.
We will assume 500pc as the distance to L977 throughout this paper.

\subsection{Derived Mass}

A precise mass determination (and accurate if not for the uncertainty
associated with the distance determination) can then be obtained for
L977 through spatial integration of the extinction map in Figure 7.
Using the standard gas--to--dust ratio (i.e., N(H + H$_2$) = 2 $\times$
10$^{21}$ cm$^{-2}$ mag$^{-1}$; Lilley 1955, Bohlin et al.  1978), and
typical mass fractions ($\sim$ 63\% H$_2$, $\sim$ 36\% He, $\sim$ 1\%
dust; e.g., Rohlfs \& Wilson 1996) we derive a mass of the surveyed
area of the cloud:  $M_{L977} = (660 \pm  30) \ (D/0.5 kpc)^2\ $ 
\msun\ where $D$ is the true distance to this molecular cloud.

\subsection{The $\sigma_{disp}$ versus A$_V$ Relation}

Our ordered map of the distribution of extinction in L977 provides
useful information about the overall large--scale structure and
distribution of mass in the cloud.  However this is done at the
expense of angular resolution.  Since each extinction measurement is
made along an individual pencil--beam through the cloud the
observations represent a random, but highly undersampled map of
extinction made at infinitely high angular resolution.  One of the most
interesting results of the first application of the NICE method to a
dark cloud was the demonstration that useful information concerning
cloud structure on small angular scales could be derived from the
randomly sampled extinction measurements.  Lada et al.  (1994) showed
that the relation between $\sigma_{disp}$, the dispersion of
extinction measurements within a square map pixel, and A$_V$, the mean
extinction derived for the map pixel, can be used to characterize
cloud structure on scales smaller than the resolution of the map
(i.e., the size of the map pixels).  In IC 5146 Lada et al.  (1994)
found that $\sigma_{disp}$ increased in a systematic fashion with
increasing A$_V$.  At the same time the dispersion in this relation
was also found to increased with A$_V$.  Through modeling, this
relation was found to be related to the form of the distribution of
the small--scale cloud structure.  In Figure 9 we present this same
relation for L977.  The same trend observed in IC 5146 is found for
L977.  A least--squares fit over the entire data set returns:

\[\sigma_{disp} \:\, = 1.93 \pm 0.11 +(0.40 \pm 0.02)\,A_V 
\,\,\,\, (mag) \]

\noindent The slope of the relation is virtually the same  as that 
found in the IC 5146 study ($0.40 \pm 0.01$). 

The increase in the value of the intercept (from 0.73 in IC 5146 to
1.93 in L977) is solely related to the higher dispersion in the colors
of background stars in the present study (Lada et al.  1994).  We
note, however, that this new data set (L977) is not as robust as the
IC 5146 data set since there are not as many measurements at high
extinction.  This is related to the fact that the total area surveyed
in the L977 study is smaller than that in IC 5146.  Apparently, the
same general behavior of the $\sigma_{disp}$ versus A$_V$ relation is
observed for L977.  As was shown by Lada et al.  (1994), this
indicates that significant structure must be present down to scales
smaller than the extinction map resolution, and moreover the dust
cannot be distributed uniformly or in discrete high--extinction clumps
on scales smaller than our resolution (0.22pc).  In a recent paper,
Padoan, Jones, \& Nordlund (1997) found that the form of the observed
$\sigma_{disp}$ versus A$_V$ relation is consistent with cloud
structure models characterized by supersonic random motions.  However,
it is not clear whether the shape of the $\sigma_{disp}$ versus A$_V$
relation is due to random spatial fluctuations in the cloud structure
or to systematic, unresolved gradients in the distribution of
extinctions (e.g., Thoraval, Boiss\'e, \& Duvert 1997).  Higher
sensitivity and angular resolution observations are needed to resolve
this issue.

\subsection{Modeling and Density Structure}

It is possible to predict the frequency distribution of detected
background stars through the target cloud as presented in Figure 6.
The decrease in the observed number of background stars for increasing
cloud depth results from the combined effect of two causes:  1)
extinction of the faintest stars in the background luminosity function
to below the limits of detection and 2) the large--scale distribution
of mass, as traced by dust extinction, inside the molecular cloud.
The existence of a rising luminosity function determines that for the
same detection limit only the intrinsically brighter (and fewer) stars
can be detected as dust extinction (e.g., A$_V$) increases.  The way
dust extinction is distributed inside the cloud will also
affect the final shape of the observed frequency distribution.

We constructed Monte Carlo models with different cloud configurations
to mimic the infrared data and predict the shape of the frequency
distribution of stars detected through the cloud.  Because of the
cloud's filamentary nature we adopted a cylindrically symmetric
synthetic cloud with a finite radius $R$.  For every model the
distribution of mass inside the model cloud was characterized by a
density profile of the form $\rho \,(r) = \rho_0 ({r\over
r_0})^{-\alpha}$ with $\alpha =$ 0, 1, 2, 3, and 4 where $r$ is the
orthogonal distance from the major axis of the cylindrical cloud.  The
density profile was integrated to produce a column density (or visual
extinction) profile for the synthetic cloud, i.e.:

\[ A_V(p) = 10^{-23} \times \rho_0r_0^\alpha \times 
\int\limits_{0}^{S} (s^2 + p^2)^{-{\alpha\over 2}}ds\  \,\,  (mag)\]

\noindent
where 

\[S = (R^2 - p^2)^{1\over2}\]
    
\noindent
and where $p$ is the projected distance of the line--of--sight from the 
major axis of the cloud, $s$ is the distance through the cloud along
the line--of--sight and $\rho_0$ and $r_0$ the appropriate scaling 
constants for the density distribution.
This column density profile was then scaled to fit the data:
the total radius of each cloud model was taken as the one that best
reproduced the observed contrast from the highest measurements of
extinction (A$_V \sim 40$ mag, at $\sim$ 40$^{\prime\prime}$ from the
center of the filament) to the lowest (1$\sigma$) measurements (A$_V
\sim 2$ mag, at $\sim$ 275$^{\prime\prime}$ from the center of the
filament).

We then constructed synthetic luminosity functions for the background
population from randomly sampling the $K$ luminosity function of the
Control Fields.  Stars from the synthetic background luminosity
function were randomly assigned a projected distance from the center
of the cloud.  This assumes a uniform spatial distribution of
background stars which is supported by the similarity between the East
and West Control Fields (see Section 3).  Next, the corresponding
extinction A$_V$, for a given density structure model, was assigned to
each star.  Noise, simulating extinction measurement errors (see next
Section) was also added to each star's extinction value.  The
extincted brightness of each star was then calculated.  Stars whose
extincted brightness was less than the limiting sensitivity of our
SQIID survey were rejected from further consideration.  One
realization was completed when the total number of extincted
detectable stars (with A$_V \le 2$) reached the observed number:  889.
To mimic the effects of the foreground stellar population we randomly
assigned a zero extinction (plus noise) to five stars (see Section
4.2) in each realization.  A total of 1000 realizations for each cloud
model ($\alpha =$ 0, 1, 2, 3, and 4) were performed.  To compare the
different model clouds and the observed frequency distribution the
resulting extinction (unbinned) distribution from each individual
realization was compared to the observed (unbinned) distribution via
the KS test.

In Figure 10 we present the comparison between the observed frequency
distribution of extinction measurements and the predictions from
clouds models with density structures having $\alpha =$ 1 (dashed
line), 2 (solid line), 3 (dotted line), and 4 (dash--dotted line).
The represented predictions correspond to the average of the 1000
realizations done for each cloud model.  Both the data and the results
from the models are binned in 2 mag wide bins.  The average KS
probability for all realizations in each model was {\it p}$_{KS}
\,(\alpha=1) \approx 10^{-3}$, {\it p}$_{KS}\, (\alpha=2) \approx
0.43$, {\it p}$_{KS} \,(\alpha=3) \approx 0.03$, and {\it p}$_{KS}
\,(\alpha=4) \approx 10^{-3}$ with rms dispersions of 10$^{-2}$, 0.24,
0.05, and 10$^{-2}$ respectively.  The constant density case,
$\alpha=0$, predicts a slowly increasing frequency distribution with a
zero probability of being, together with the observed distribution,
randomly drawn from the same parent distribution.  This model cloud is
not represented on Figure 10 for the sake of clarity.  Our Monte Carlo
simulation suggests that the large--scale distribution of matter
inside L977 is most consistent with a density structure $\rho \,(r)
\propto r^{-2}$.  We note that an isothermal cylinder has a density
structure $\rho \,(r) \propto r^{-4}$ (Ostriker 1964).

In Figure 11 we present the averaged extinction profile of the southern
portion of L977 where the center to edge contrast in extinction is the
highest.  Each point in this figure represents the average extinction
along the North--South direction of 7 adjacent extinction map pixels
covering a declination of $\sim 5$ arcmin on the southern part of the
NIR survey, where the filament is roughly aligned to the North-South
direction.  The error bars represent the rms dispersions in the
average of the pixels.  The ``bump'' around 300 arcsec is caused by a
small secondary condensation within L977.  The dotted lines are column
density profiles corresponding to radial density laws $\rho \,(r)
\propto r^{-\alpha}$, for $\alpha=$0, 1, 2, and 3.  The best fit
between the observed radial profile and the various model profiles
occurs for a $\rho \,(r) \propto r^{-2}$ density structure.  The
1$\sigma$ uncertainty in the averaged extinction measurements is now
A$_V\sim 0.3$ mag.  We conclude from our Monte Carlo simulation and
from the observed radial profile of the cloud that the distribution of
matter inside L977 between 2 $<$ A$_V$ $<$ 40 mag (or 1 $<$ $r$ $<$
0.1pc) is consistent with a density profile of $\rho \,(r) \propto
r^{-2}$.

We do not expect the density in the inner part of L977 to reach
infinity as $r \rightarrow 0$.  Clearly a break at some small $r_0$
($r_0$\ltsima 0.1pc; from Figure 11) must occur.  With the present data
we cannot probe the very inner part of L977 because we do not detect
stars through it.  In fact, even with new and more sensitive
near--infrared arrays this will not be an easy task since the
probability of having a fairly bright background star within a 
smaller solid angle (the denser regions of the cloud) is also
increasingly small.

\subsection{The Effect of Galactic Latitude on the NICE Method}

It was a goal of this article to analyze the effects of a richer but
more complicated background of field stars --- at lower galactic
latitude --- on the application of the NICE method.  An extensive
discussion on the limitations and uncertainties in this method is
given by Lada et al.  (1994).  Here we will only be concerned about
the limitations that result from applying the NICE method to low
Galactic latitude molecular clouds by comparing the previous study
(IC 5146, $b = -5$\deg) with the present one ($b = 2$\deg).

The angular resolution of the dust extinction map is limited by the
surface density of stars detected through the target cloud at both the
$H$ and $K-$bands.  For the same detection limit it is clear that
better resolution, hence more complete information on the small--scale
spatial distribution of dust, can be obtained for clouds that lie in
the foreground of rich star fields close to the Galactic plane.  On
the other hand it was unclear before the present study what was the
contribution of distant luminous giant stars, suffering extraneous
extinction, to the increase in the dispersion of the mean ($H-K$)
color of control field stars at these low Galactic latitude fields.
The NICE method is sensitive to this mean color since it assumes that
the intrinsic $(H-K)$ color of stars behind the cloud is accurately
represented by the mean color of the control fields.  A successful
application of the method requires, to first order, the dispersion on
the mean ($H-K$) color for the control fields to be significantly
smaller than the observed range in $(H-K)$ colors for stars seen
through the target cloud.

A direct consequence of observing at low Galactic latitudes is an
appreciable increase in the mean and dispersion of the observed
$(H-K)$ (and ($J-H)$) color of stars:

\[<(H-K)>_{IC5146} \:\, = 0.13 \pm 0.08 \:\, mag \,\,\,\,\,\,\,\, (b =
-5^\circ ) \]

\[<(H-K)>_{L977} \:\, = 0.20 \pm 0.13 \:\, mag \,\,\,\,\,\,\,\,\, (b =
2^\circ ) \]

\noindent While the value of the mean observed $(H-K)$ color will
function as a zero--point for the derived extinction measurements with
the NICE method, the dispersion in this value ---
$\sigma(H-K)_{intrinsic}$ --- sets the threshold (1$\sigma$ confidence
level) above which extinctions toward background stars can be reliably
measured.  The observed increase in dispersion, caused by distant
reddened giants detected at low Galactic latitude, is going to raise
this threshold.

For the IC 5146 study the threshold value was $\sim$ 1 magnitude of
visual extinction and it was similar to the measurement uncertainty on
a pixel of the derived extinction map:

\[ \sigma_{pixel}\ =\ {15.9/N} \sqrt{\Sigma_{i=1}^N (\sigma(H-K)^2_i +
\sigma(H-K)^2_{intrinsic})} \,\,\,\,\,\,\,\, (mag)\]

\noindent 
where $\sigma(H-K)_i$ is the photometric error in the
derived color for a star in a given map pixel and N is the number of
stars in that pixel.  For the L977 study the larger value of
$\sigma(H-K)_{intrinsic}$ raises the threshold to $\sim$ 2 magnitudes
of visual extinction while the uncertainty in the extinction map is 
still $\sim$ 1 magnitude of visual  extinction.  The uncertainties 
on a map pixel are similar in both studies because the average number 
of stars inside each pixel map is now 8 (instead of 5 for the IC 5146 
study) which compensates for the larger $\sigma(H-K)_{intrinsic}$ 
observed in the present study.

We conclude that the consequence of applying the NICE method at low
Galactic latitudes is an increase of as much as a factor of 2 in the
1$\sigma$ confidence level in the derived extinction measurements
toward background stars.  This is directly related to the detection at
low galactic latitudes of a large number of giant stars along the
plane of the Galaxy affected by random, unrelated, amounts of
extinction and affects somewhat the method's aptitude to address
small--scale cloud structure (e.g., the increase in random dispersion
around the $\sigma_{disp}$ versus A$_V$ relation).  Nonetheless, and
because of the comparatively higher number of stars detected at these
very low galactic latitudes $(b = 2^\circ)$ \footnote {The warping of
the galactic disk makes this latitude essentially $\sim 0^\circ$.},
the capacity of the method to study the large--scale distribution of
dust in a molecular cloud is essentially the same as an application of
the method at the same galactic longitude but at a galactic latitude
$(b = -5^\circ)$ free from ``contamination'' of distant giant stars.
The NICE method remains a powerful method for mapping the distribution
of dust through a molecular cloud.

\section{Summary}

We applied the NICE method to the L977 molecular cloud.
The main findings of this paper are as follow:

\begin{itemize}

\item We use measurements of the near--infrared color excess and
positions of the 1628 brightest stars in our survey to directly
measure dust extinction through the cloud following the method
described by Lada et al.  (1994).

\item We construct a detailed map of the distribution of dust
extinction of L977 molecular cloud at an effective angular resolution
of 90$^{\prime\prime}$ by spatially convolving the individual
extinction measurements.  We derive a distance of 500 $\pm$ 100
pc towards this cloud via a comparison of source counts with
predictions of a galactic model.  From integration of the extinction
map we determine the total mass of the cloud to be $M_{L977} = (660
\pm 30) \ (D/500 pc)^2$ \msun where $D$ is the true distance to the
cloud.

\item We find a correlation between the measured dispersion
in our extinction determinations and the extinction which is very
similar to that found for the dark cloud IC 5146 in a previous study.
We interpret this as evidence for the presence of structure on scales
smaller than the 90$^{\prime\prime}$ resolution of our extinction map.

\item To further investigate the structure of the cloud we construct
the frequency distribution of the 1628 individual extinction
measurements in the L977 cloud.  The shape of the distribution is
similar to that of the IC 5146 cloud suggesting that both
clouds are characterized by a similar structure.  Monte Carlo modeling
of this distribution suggests that between 2 $<$ A$_V$ $<$ 40 mag (or
roughly 1 $<$ $r$ $<$ 0.1 pc) the material inside L977 is
characterized by a density profile $\rho \,(r) \propto r^{-2}$.
Direct measurement of the radial profile of a portion of the cloud
confirm this result.

\item From comparison between the colors of stars seen through L977
and color of stars in the control fields we derive the slope of the
near--infrared extinction law to be $E(J-H)/E(H-K) = 1.6 \pm 0.1$.

\item At the lower galactic latitude of L977, we find both the mean
and dispersion of the infrared colors of field stars to be larger than
observed toward IC 5146.  This produces an increase of about a factor
of 2 in the minimum or threshold value of extinction that can be
reliably measured toward L977 with this technique.  Nevertheless the
accuracy in an extinction map pixel is not significantly different
toward L977 due to the increased number of field stars at this
latitude.  We also find an increase in the number of detected giant
stars at the low galactic latitude of the survey by almost a factor of
two.  Most of these excess stars suffer extraneous extinction and are
probably red giants seen along the disc of the Milky Way up to
distances $\sim$ 15 kpc and reddened by unrelated background molecular
clouds along this direction of the Galaxy.  We discuss a possible
application of this observable to galactic structure studies on the
plane of the Galaxy.

\end{itemize}

\acknowledgments

This work improved from helpful discussions with Maria Lu\ii sa
Almeida, Alyssa Goodman, Diego Mardones, and August Muench.  This
study was supported by the Smithsonian Institution Scholarly Studies
Program SS218--3--95.  JA acknowledges support from the Funda\cao\
para a Ci\^encia e Tecnologia (FCT) Programa Praxis XXI, graduate
fellowship BD/3896/94, Portugal.

\newpage
\appendix
\section{The Effect of Galactic Latitude on NIR Color--Color
Diagrams}

A comparison between L977 control fields ($b = $ 2\deg) with control
fields at essentially the same Galactic longitude but not as close to
the Galactic plane (IC 5146 molecular cloud, $b = -$5\deg, Lada et al.
1994) reveals significant differences.  In Figure A1 we present the
color--color diagrams for both regions.  The contours represent
surface density of stars in the [($J-H), (H-K)$] color--color plane.
Contours are 30\% (2$\sigma$), 45\%, 60\%, 75\%, and 90\% of the
maximum number of stars/mag$^2$ and were obtained by sampling the
color--color plane at Nyquist frequency with a square filter with a
resolution equal to the average photometric error (0.1 mag) in both
samples.  The resolution is represented by the square box at the upper
right corner of both diagrams.  The dwarf and giant sequence (solid
line) and the reddening vector (dashed line) are as in Figure 3.  Both
IC 5146 and L977 control fields regions show a similar, but not
identical, bimodal distribution of sources on the color--color plane,
i.e., a blue group centered at the color of a mid G main--sequence
star and a red group centered at the color of an early K
giant/main--sequence star.  For the SQIID detection limit the blue
group is dominated by F main-sequence stars located at distances
smaller than 1 kpc while the red group is dominated by early K giant
stars located at a broad range of distances that can surpass 15 kpc
(Alves 1998).

Within the sampling resolution the blue group seen in the L977 control
fields is similar to the one in IC 5146 control fields.  This is not
surprising since both groups likely consist of a similar population of
{\it local} main-sequence stars.  If relevant at all, the somewhat
higher scatter observed in the L977 blue group can be due to higher
interstellar extinction towards this direction which lies closer to
the plane of the Galaxy.  A significant difference in both diagrams is
revealed in the comparison of the red groups.  The center of the L977
red group is not only displaced along the reddening vector when
compared to IC 5146 red group but its overall shape is changed:  the
uniform distribution of stars around a preferred point in the
color--color plane for the IC 5146 red group is replaced by an
elongated distribution along the reddening vector for the L977 red
group, indicating the presence of additional differential extinction
for this population of stars.  Also, the fraction of sources belonging
to the red group (essentially the ratio [\#giants/\#main--sequence
stars]) detected in the IC 5146 control fields is 37\% while for L977
this fraction rises to 56\%, almost a factor of two higher.

There is a clear ``contamination'' by reddened background giants in
our $JHK$ color--color diagram closer to the Galactic plane.  Giant
stars are easily detected even at very large distances because:  1)
they are luminous NIR emitters (M$_K \sim -3$ mag and T $\sim$ 4000 K
for an early K giant), 2) the opacity of the plane of the Galaxy at
these wavelengths (1.25 to 2.2 $\mu$m) is modest, and 3) being the
evolved counterparts of Sun--like stars, they are abundant.  These
facts alone do not explain the observed differences for the two
fields.  However, for low Galactic latitudes (as it is the case for
the L977 control fields) we are basically observing along the Galactic
disk virtually detecting giants to its edge.  This can explain the
observed increase in the ratio [\#giants/\#main--sequence stars] ---
this increase is in itself a measure of the relative distribution
perpendicular to the plane of the giant population.  Moreover, the
farther a detected giant is along the plane the more likely it is to
be affected by extinction.  This is most probably the cause of the
displacement and scatter along the reddening vector for the red group
in the L977 control fields color--color diagram in Figure A1.  The
similarity between L977 control fields East and West (Figure 3),
separated by 1\deg\ in the sky, supports this idea by suggesting that
the observed effect is characteristic of this part of the Milky Way.

Although the increase in the dispersion of the mean color of
background stars is a nuisance in the present study there are obvious
fruitful applications of this quantity to galactic structure studies
along the plane of the galaxy.  Classical galactic structure studies
have always been severely hampered and confined to a few kpc around
the Sun by the high opacity of the galactic plane that concealed from
direct observation the regions where nearly all the luminous galactic
mass is located.  We note that the NIR is the wavelength of choice for
this kind of study because of its comparative insensitivity to:  1)
dust extinction, and 2) dust emission that begins to dominate at
wavelengths longer than 5 $\mu$m.  Moreover, the $JHK$ color--color
diagram is a good discriminator of distant giant stars at low Galactic
latitudes, as it is shown in Figure A1, hence a powerful tool to address
galactic structure questions at the optically inaccessible low galactic 
latitudes, e.g., the large scale distribution of dust along the galactic
plane, or the structure of the Galaxy's outer disk.

\begin{figure}
\figurenum{1}
\plotfiddle{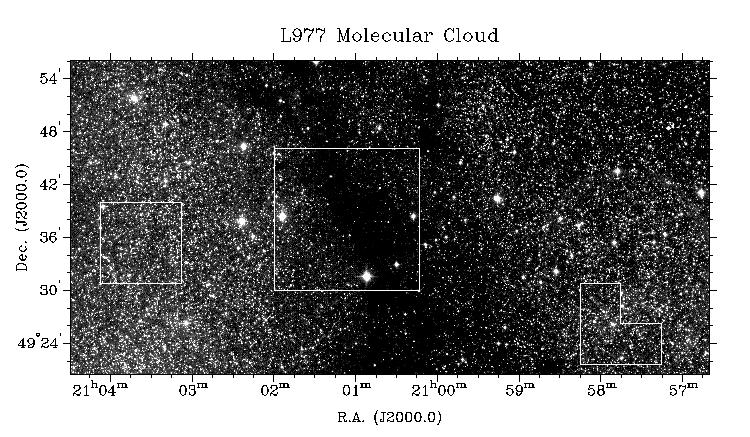}{14cm}{-90}{80}{80}{-310}{400}
\caption{Finding chart for L977.  The area surveyed in
the 3 NIR bands is represented as the central square overlaid on the
Digitized Sky Survey red POSS plate.  The L977 molecular cloud is
readily seen as the zone of obscuration inside this square and seen
against the rich star field that characterizes this region of the
galaxy.  The boxes to the East and to the West of L977 define the
regions where the East and West control fields were taken.  They are
spatially separated in the sky by 1 degree  ($\Delta l = 0.7$\deg, and
$\Delta b = 0.8$\deg).}
\end{figure}

\begin{figure}
\figurenum{2}
\plotfiddle{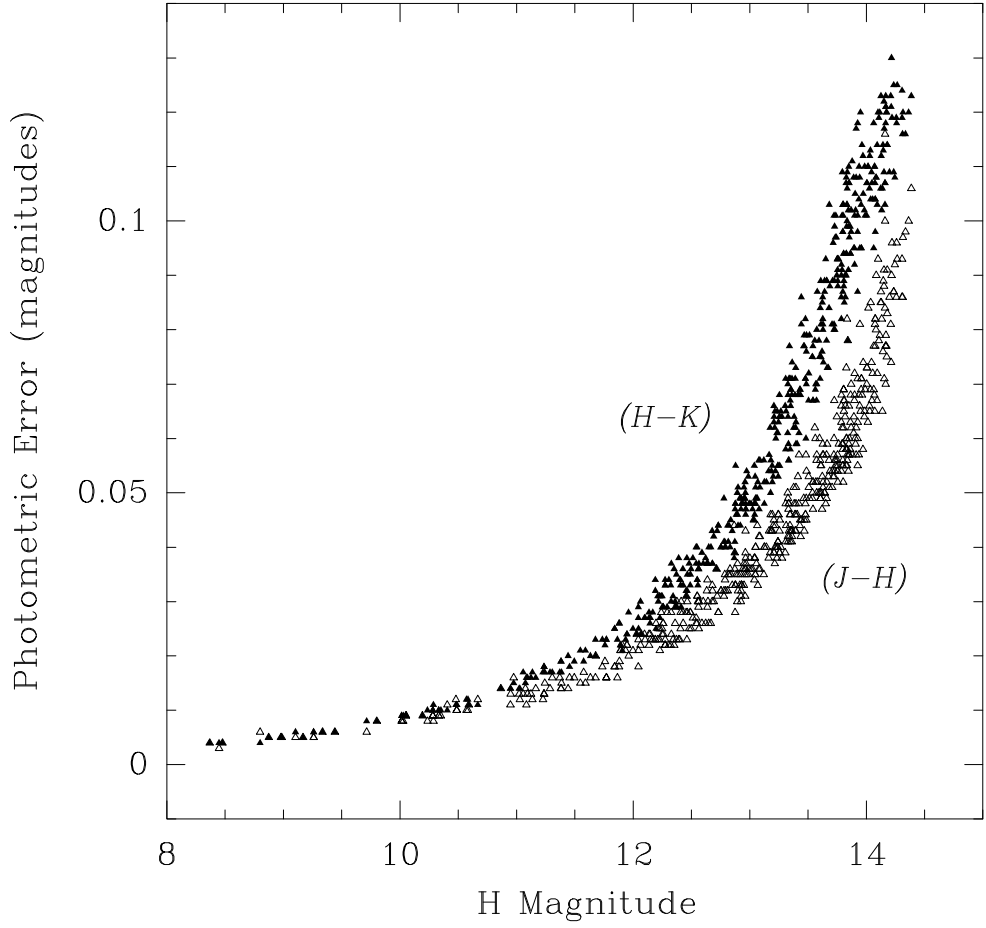}{12cm}{0}{80}{80}{-250}{0}
\caption{The photometric errors in the colors ($(J-H)$
open triangles; $(H-K)$ filled triangles) as a function of magnitude
$(H)$ for stars observed toward L977.}
\end{figure}

\begin{figure}
\figurenum{3}
\plotfiddle{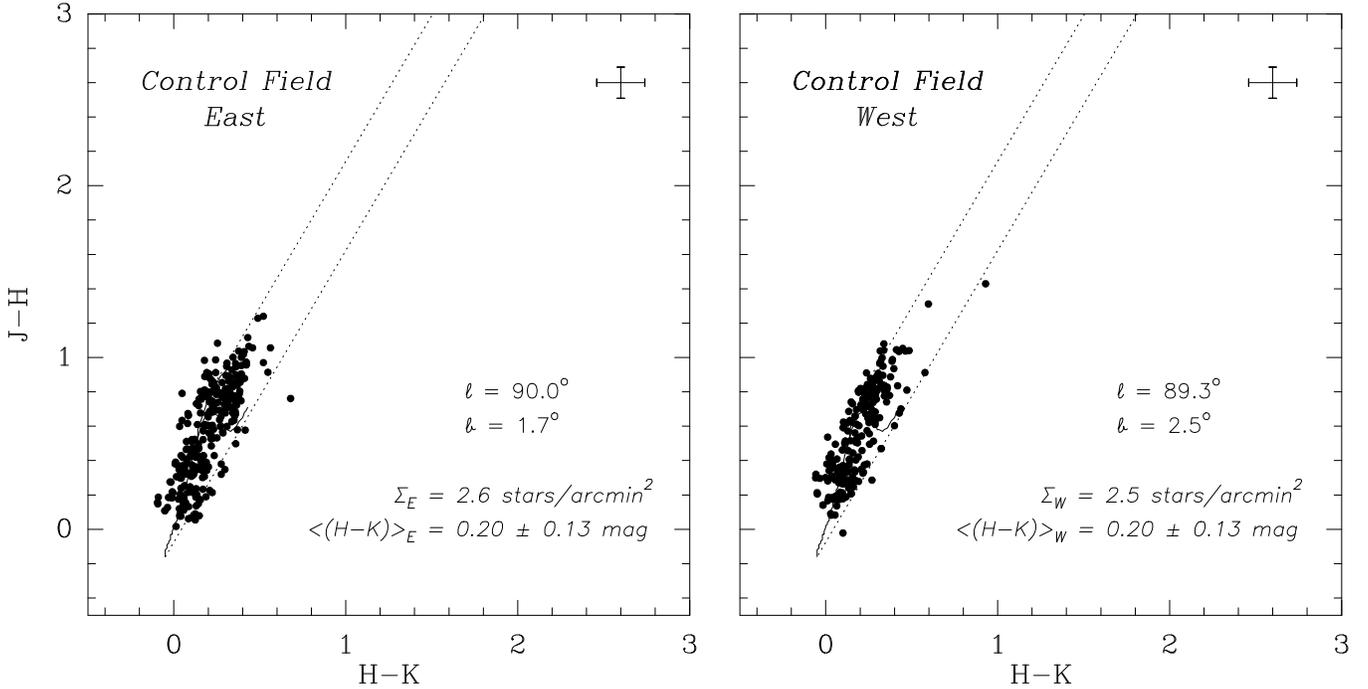}{14cm}{-90}{80}{80}{-310}{400}
\caption{Near--infrared ($JHK$) color--color diagrams for
East and West control field regions.  Also plotted as a solid line on
both color--color diagrams are the locus of points corresponding to
unreddened main sequence and giant stars (Koornneef 1983).  The two
parallel dashed lines define the reddening band for both main sequence
and giant stars and were taken from Rieke \& Lebofsky (1985).  The
typical photometric error for a 14$^{th}$ $K$ magnitude star ---
roughly a 10 sigma detection --- is displayed in the upper corner of
both diagrams.  The surface density of detected sources in the
$JHK-$bands for the East control field region ($\Sigma_E$) is 4\%
higher than in the West control field region ($\Sigma_W$) --- 2.6
against 2.5 stars per square arcmin --- most likely due to the fact
that the East control field region is almost 1\deg\ closer to the
Galactic plane.  The mean $(H-K)$ and $(J-H)$ colors and dispersions
for the two regions are, nevertheless, virtually the same.  See text
for full discussion.}
\end{figure}

\begin{figure}
\figurenum{4}
\plotfiddle{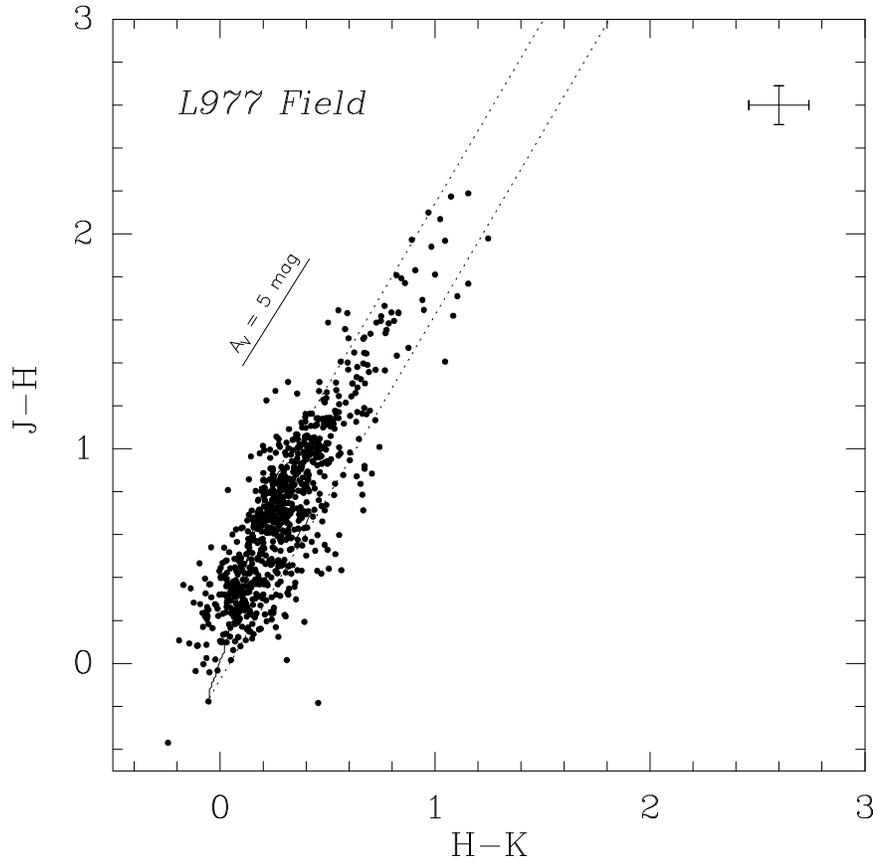}{14cm}{-90}{100}{100}{-400}{500}
\caption{Near--infrared ($JHK$) color--color diagram for
the L977 field defined in Figure 1 by the central square region.  The
main sequence and giant locus, the reddening vector, as well as the
photometric errors are the same as in Figure 2.  The displacement in
this diagram caused by 5 magnitudes of extinction is represented by
the solid line to the left of the reddening vector.  The ten most
deviant sources scattered around the reddening vector were all found
to be very close to the 10 sigma detection limit.}
\end{figure}

\begin{figure}
\figurenum{5}
\plotfiddle{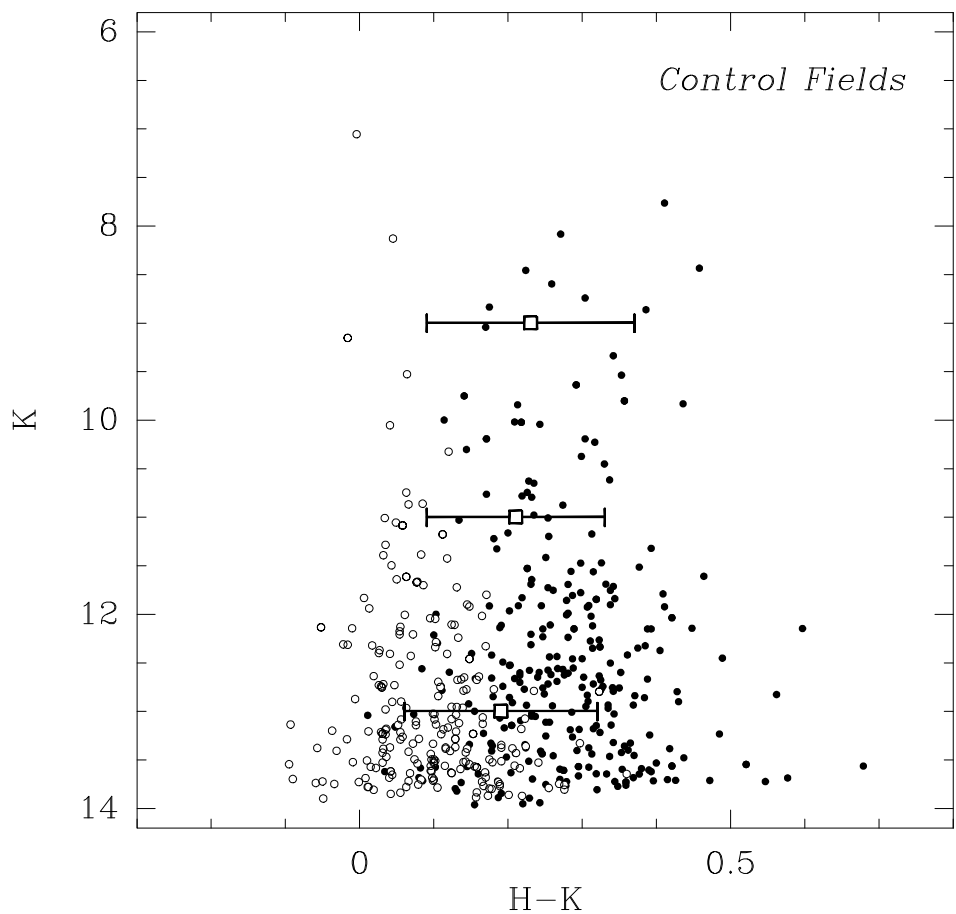}{14cm}{-90}{80}{80}{-310}{450}
\caption{Color--magnitude diagram for all the 522 stars
in the control field regions.  The open circles represent sources with
$(J-K) < 0.5$ mag (the blue group, described in Sec.  3.1) and the
filled circles represent sources with $(J-K) > 0.5$ mag (the red
group).  The open squares and the error bars represent the mean and
the dispersion, respectively, of the $(H-K)$ color in the (2 magnitudes
wide) bins centered at $K =$ 9, 11, and 13.}
\end{figure}

\begin{figure}
\figurenum{6}
\plotfiddle{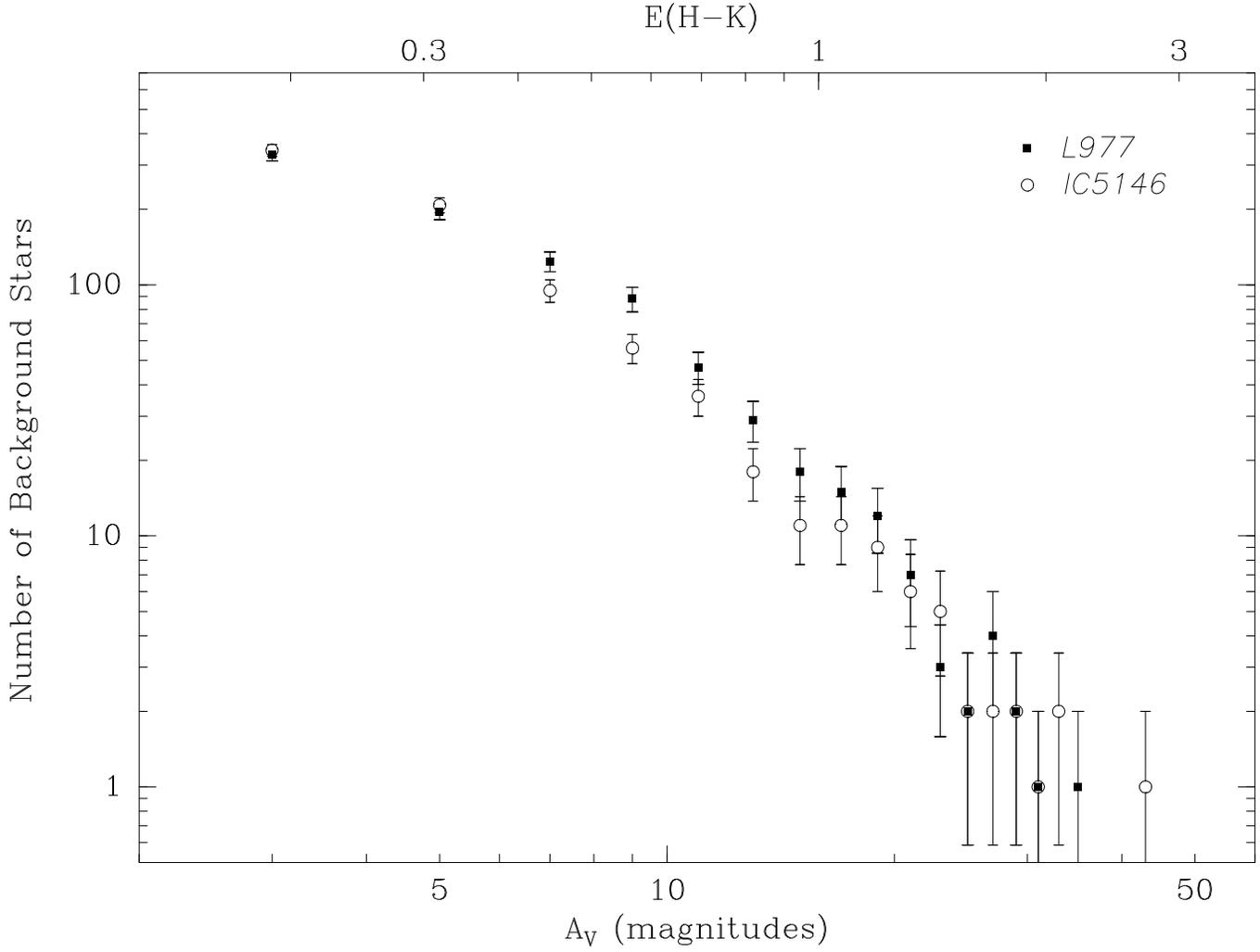}{14cm}{-90}{80}{80}{-310}{450}
\caption{Frequency distribution of detected background
stars through L977 and IC 5146 molecular cloud. The similarity
in the shape of both distributions suggests a similar physical
structure for these two clouds.}
\end{figure}

\begin{figure}
\figurenum{7}
\plotfiddle{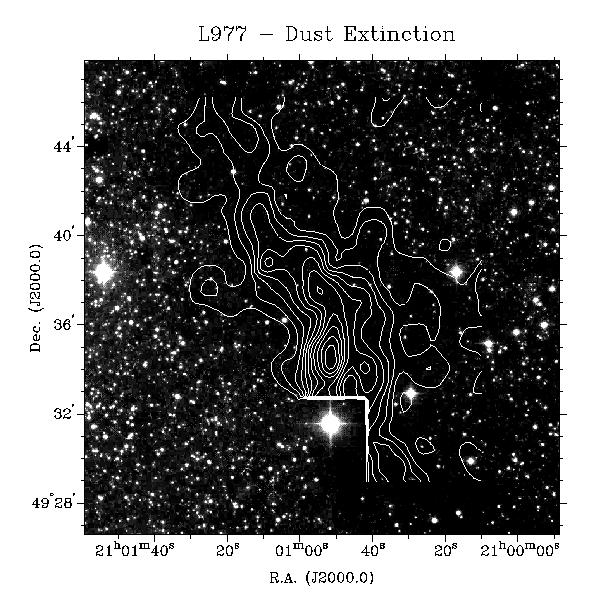}{14cm}{-90}{80}{80}{-310}{460}
\caption{L977 dust extinction map derived from the
infrared $(H-K)$ observations.  Contours start at 4 visual magnitudes
of extinction (4$\sigma$) and increase in steps of 2 magnitudes up to
24 magnitudes of visual extinction. The effective resolution of the map
is 90$^{\prime\prime}$ (0.22pc at an estimated distance of 500 pc).
The contours are overlaid on the digitized POSS red plate.  The
overall shape of the map correlates well with the shape of the more
opaque regions of the molecular cloud.  The discontinuity near the
bright star SAO 050355 (a M3 star) is due to lack of one SQIID field 
in our data, not taken to avoid potential detector damage.}
\end{figure}

\begin{figure}
\figurenum{8}
\plotfiddle{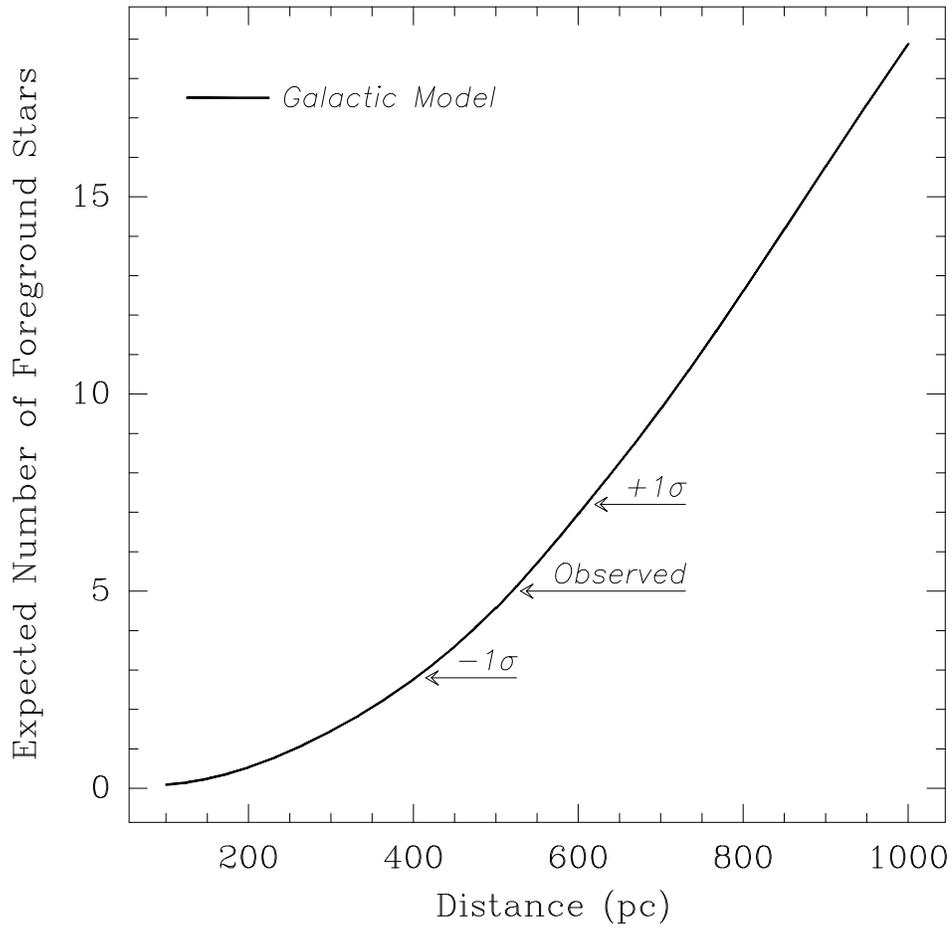}{14cm}{-90}{80}{80}{-310}{450}
\caption{Expected number of foreground stars toward the
denser regions ($\sim$ 36 arcmin$^2$) of L977 as a function of
distance.  Five foreground stars were identified in the NIR survey
toward these regions.}
\end{figure}

\begin{figure}
\figurenum{9}
\plotfiddle{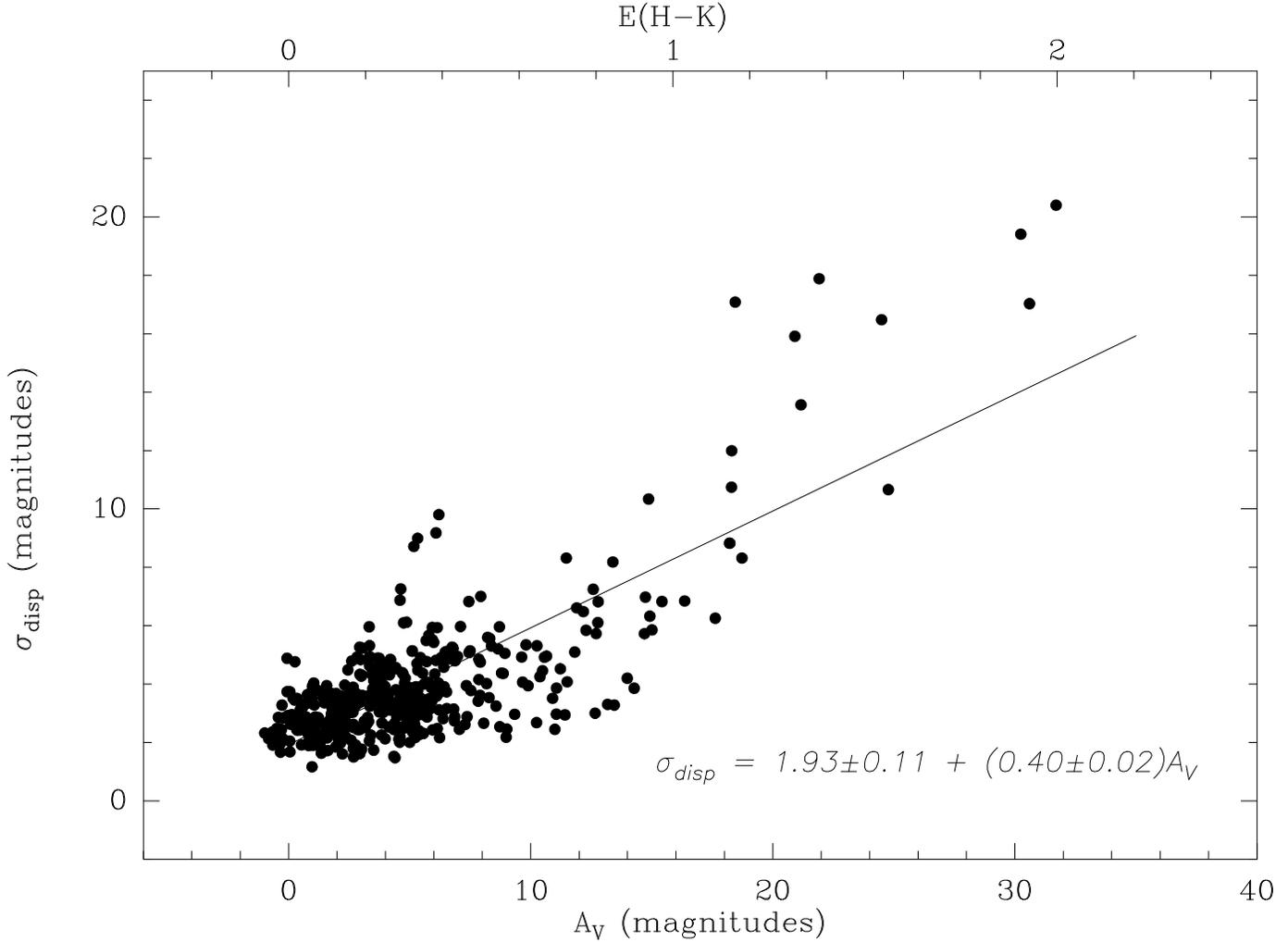}{14cm}{-90}{80}{80}{-310}{450}
\caption{The relation between $\sigma_{disp}$, the
dispersion in the extinction measurements, with visual extinction
A$_V$.  Also plotted is the least--squares linear fit to the data over
the entire range of extinction.  The same general behaviour is
observed in the IC 5146 study (Lada et al.  1994).}
\end{figure}

\begin{figure}
\figurenum{10}
\plotfiddle{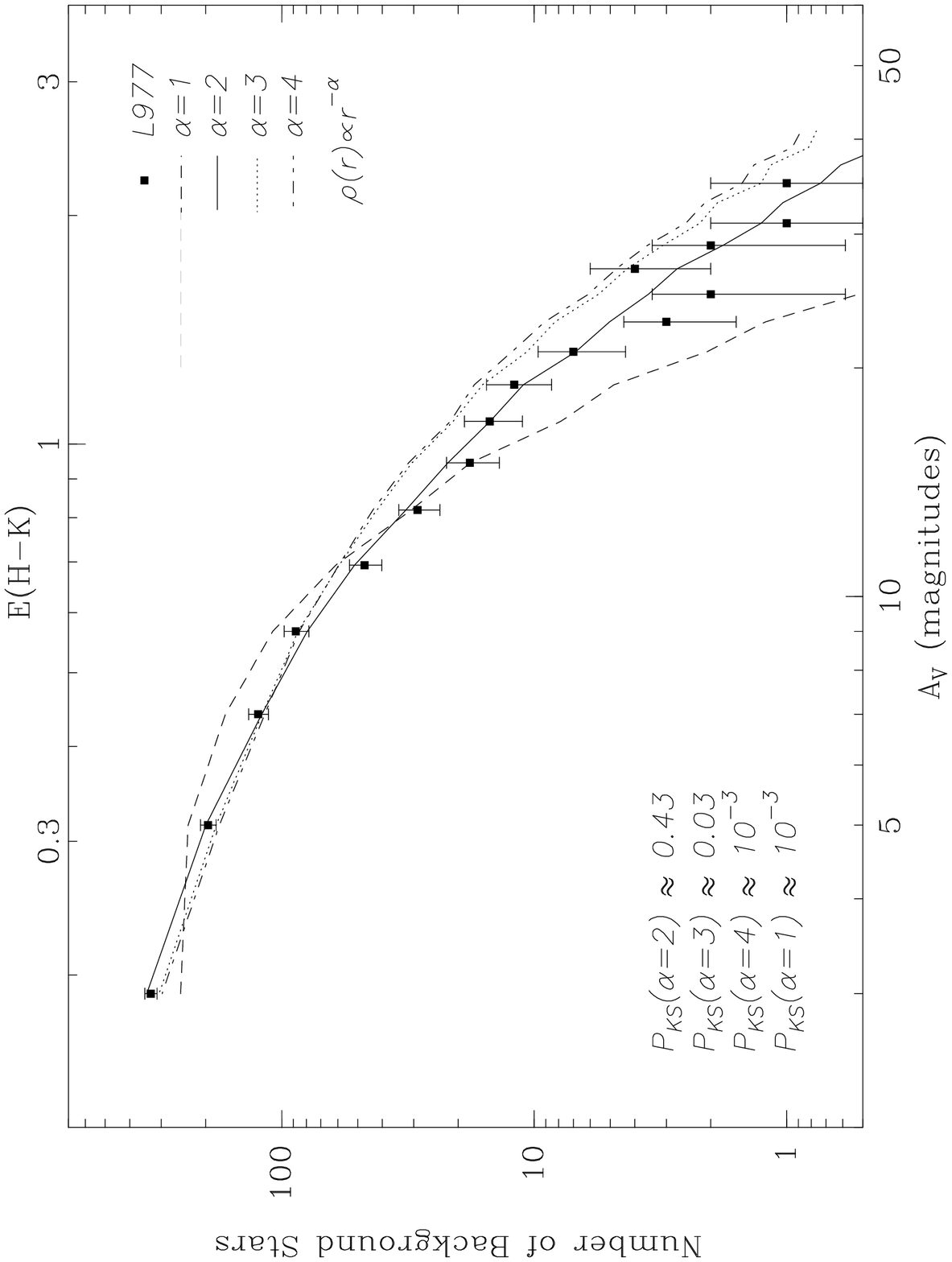}{14cm}{-90}{80}{80}{-310}{450}
\caption{Observed frequency distribution of extinction
measurements for L977 and the predictions from clouds models with
density structures $\rho \,(r) \propto r^{-\alpha}$ having $\alpha =$
1 (dashed line), 2 (solid line), 3 (dotted line), and 4 (dashed--dotted
line). The represented predictions correspond to the average of the 1000 
realizations done for each cloud model. The average KS probability for all
realizations in each model is presented.}
\end{figure}

\begin{figure}
\figurenum{11}
\plotfiddle{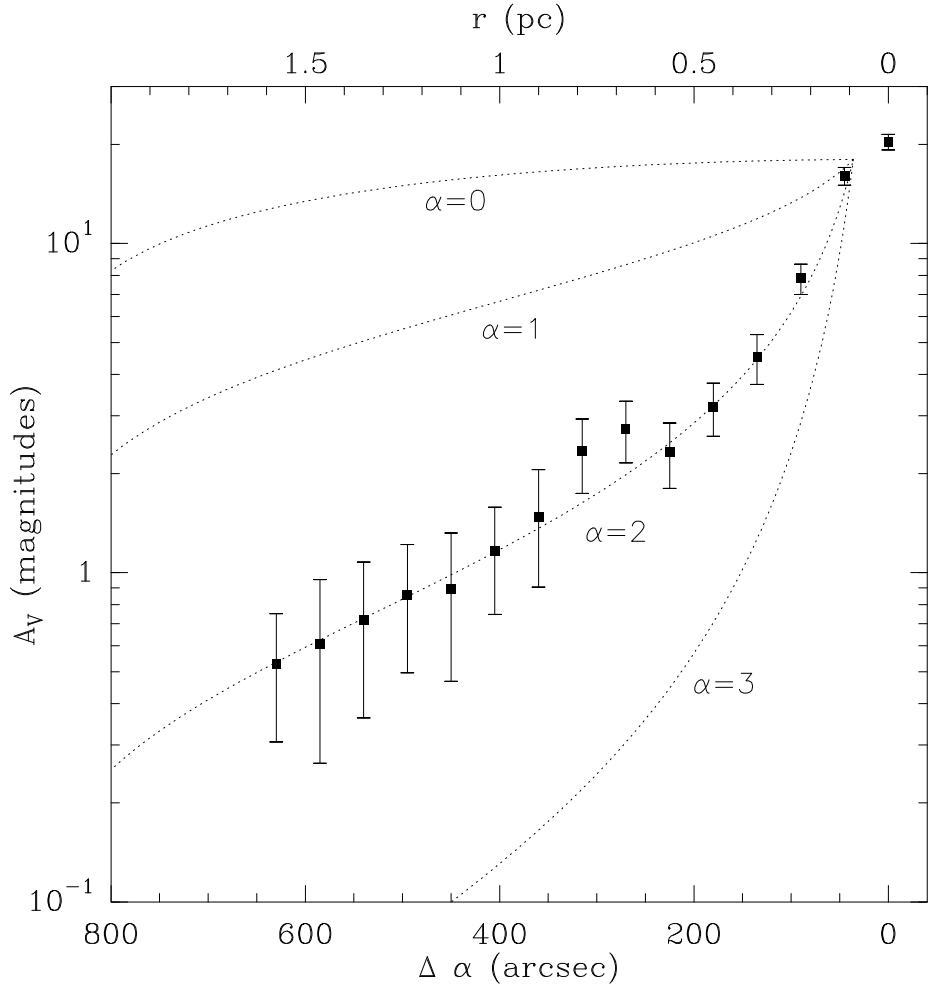}{14cm}{-90}{80}{80}{-310}{450}
\caption{Averaged radial profile of L977.  The ``bump''
around 300 arcsec is caused by a small secondary condensation within L977.  
The dotted lines correspond to radial density profiles $\rho \,(r) \propto
r^{-\alpha}$, for $\alpha=$0, 1, 2, and 3. The upper distance scale
assumes a distance of 500pc to L977.}
\end{figure}

\begin{figure}
\figurenum{A1}
\plotfiddle{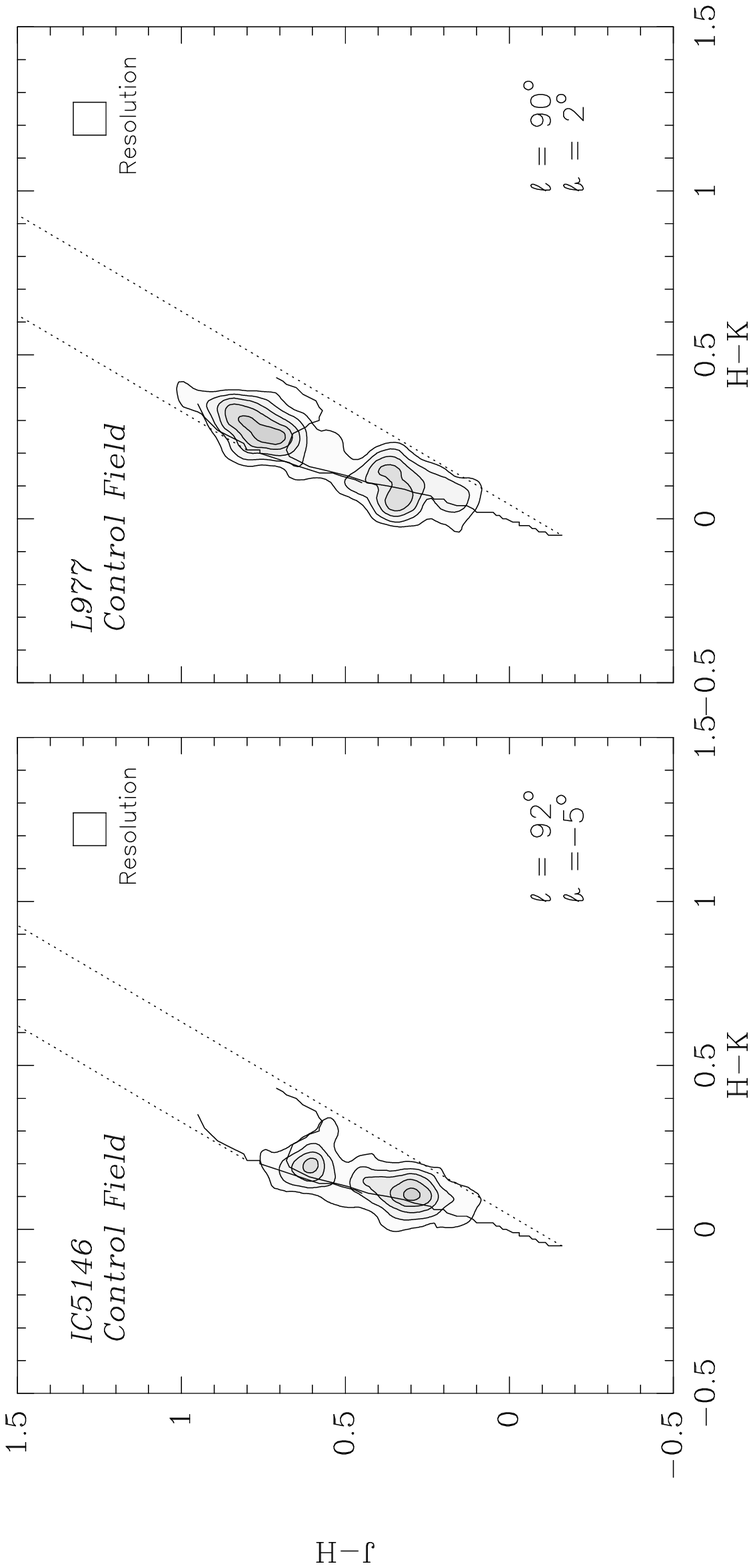}{14cm}{-90}{80}{80}{-310}{400}
\caption{a) and b) Near--infrared ($JHK$) color--color
diagrams for IC 5146 control fields (IC 5146 molecular cloud, $b =
$-5\deg, Lada et al.  1994) and the L977 control fields ($b = $
2\deg).  The contours represent surface density of stars in the
[($J-H), (H-K)$] color--color plane.  Contours are 30\% (2$\sigma$),
45\%, 60\%, 75\%, and 90\% of the maximum number of stars/mag$^2$ and
were obtained by sampling the color--color plane at Nyquist frequency
with a square filter with a resolution equal to the average
photometric error in both samples (0.1 mag).  The dwarf and giant
sequence (solid line) and the reddening vector (dashed line) are as in
Figure 2.  Note the effects of the Galactic plane on the bimodal
distribution of sources in the color--color plane.}
\end{figure}

\end{document}